\documentclass[12pt]{article}
 \usepackage{a4wide}
\usepackage{authblk}

\usepackage[breaklinks]{hyperref}
\usepackage{booktabs}
\usepackage[superscript]{cite}
\newcommand{\refcite}[1]{[\citen{#1}]}
\usepackage{mathptmx} 
\usepackage{graphicx} 
\usepackage[utf8]{inputenc}
\usepackage{amsmath,amssymb,amsthm}
\usepackage{leftidx,bbding,extarrows}
\usepackage{tikz,tikz-3dplot}
\usetikzlibrary{matrix,arrows,shapes,decorations.pathmorphing}
 
\usepackage[mathscr]{eucal}
\newcommand{\Mbb}{{\boldsymbol{\EuScript{M}}}}

\DeclareMathAlphabet{\mathcal}{OMS}{cmsy}{m}{n}

\makeatletter
\g@addto@macro\bfseries{\boldmath}
\makeatother

\DeclareMathAlphabet\BEuScript{U}{eus}{b}{n}

\newcommand{\CoinX}[1]{C_0^\infty({#1})}

\newcommand{\kin}{\text{kin}}

\newcommand{\Af}{{\mathscr{A}}}
\newcommand{\Lf}{{\mathfrak{L}}}
\newcommand{\Tf}{{\mathfrak{T}}}

\newcommand{\ogth}{{\mathfrak o}}
\newcommand{\tgth}{{\mathfrak t}}

\newcommand{\supp}{{\rm supp}\,}

\newcommand{\CC}{{\mathbb C}}
\newcommand{\RR}{{\mathbb R}}

\newcommand{\HH}{{\mathscr H}}

\newcommand{\Gc}{{\mathcal{G}}}
\newcommand{\Lc}{{\mathcal{L}}}
\newcommand{\Mc}{{\mathcal{M}}}

\newcommand{\Zc}{{\mathcal{Z}}}

\usepackage{bm}
\DeclareMathAlphabet{\mathbfsf}{\encodingdefault}{\sfdefault}{bx}{n}

\DeclareBoldMathCommand{\Cb}{{\boldsymbol{C}}}
\DeclareBoldMathCommand{\Db}{{\boldsymbol{D}}}
\DeclareBoldMathCommand{\Lb}{{\boldsymbol{L}}}
\DeclareBoldMathCommand{\Mb}{{\boldsymbol{M}}}
\DeclareBoldMathCommand{\Nb}{{\boldsymbol{N}}}

\newcommand{\nto}{\stackrel{.}{\to}}

\DeclareMathOperator{\ad}{ad}

\DeclareMathOperator{\SL}{SL}

\newcommand{\ip}[2]{\langle #1\vert#2\rangle}
\newcommand{\id}{{\rm id}}
\newcommand{\II}{{\mathbf{1}}}

\newcommand{\Ts}{{\sf T}}

\newcommand{\FLoc}{{\sf FLoc}}
\newcommand{\Loc}{{\sf Loc}}

\newcommand{\Set}{{\sf Set}}

\newcommand{\Alg}{{\sf Alg}}

\newcommand{\Phys}{{\sf Phys}}

\newcommand{\Ac}{{\mathcal A}}

\newcommand{\Bf}{{\mathscr B}}

\newcommand{\Df}{{\mathscr D}}

\newcommand{\Sf}{{\mathcal S}}

\DeclareMathOperator{\Fld}{Fld}

\newcommand{\dvol}{d{\rm vol}}

\newcommand{\LL}{\mathcal{L}}

\newcommand{\rce}{\text{rce}}

\def\beq{\begin{equation}}
\def\eeq{\end{equation}}

\newtheorem{theorem}{Theorem}[section]

\newtheorem{definition}[theorem]{Definition}

\newtheorem{axiom}[theorem]{Assumption}

\begin{document}

\title{Locally covariant quantum field theory \\ and the spin-statistics connection}

\author[1]{Christopher J. Fewster\thanks{\tt chris.fewster@york.ac.uk}}
\affil{Department of Mathematics,
University of York, Heslington, York YO10 5DD, United Kingdom.}

\date{February 5, 2016}
\maketitle

\begin{abstract}
The framework of locally covariant quantum field theory, an axiomatic approach
to quantum field theory in curved spacetime, is reviewed. As a specific focus,
the connection between spin and statistics is examined in this context. A new approach is given, which allows for a more operational description of theories with spin and for the derivation of a more general version of the spin-statistics connection in curved spacetimes than previously available. This part of the text is based on arXiv:1503.05797 and a forthcoming publication; the emphasis here is on the fundamental ideas and motivation.
\end{abstract}



\section{Introduction}

The purpose of this contribution is two-fold: on one hand, it serves as
a summary of my talk in the QF2 session on the spin-statistics connection (section~\ref{sec:spinstats});
on the other, at the request of the session organisers, it provides an
expository review of locally covariant quantum field theory in curved spacetimes (QFT in CST)
(section~\ref{sec:LCQFT}). 

In the context of a Marcel Grossmann meeting, there should be no need to justify the study of QFT in curved spacetimes. However, it is worth emphasising that locally covariant QFT has two
main differences from usual practice of QFT in CST: it is an axiomatic approach, and it aims to discuss arbitrary spacetime backgrounds, rather than specific examples. The motivation for the latter has several aspects.
First, one wishes to gain a perspective that
is independent of special features of particular spacetimes, but democratically
implements the same physics (in some sense) in all of them.  This is motivated by
the practical reason that the spacetime we inhabit does not exhibit any symmetries on 
small scales, but seems to be well-approximated on large scales by spacetimes that do,
and a setting in which such approximations can be controlled is desirable. 
Second, allowing for arbitrary backgrounds gives one flexibility to 
model macroscopic material features (e.g., stars or apparatus etc)
by the geometry of the background rather than as complicated
configurations of a QFT. Third, to embed the principle of locality from the start,
it is expedient to seek a framework in which the formulation in a given spacetime region
is (in a suitable sense) independent of the geometry in its causal complement. 

Axiomatic approaches to QFT have been developed since the 1950's. They
arose from concerns about the mathematical deficiencies of QFT at that time,
with the aim to `kill it or cure it'.\cite{StreaterWightman} At a basic level, 
the goal is to write down precisely what a quantum field theory
aspires to be, to draw out the general consequences (e.g., a spin-statistics
connection, PCT theorem, or no-go results like Haag's theorem) that follow from them and thereby to 
provide guidance for attempts to rigorously construct models of QFT.  
It is striking that, despite the undoubted successes of QFT, the
mathematical status of (nonperturbative) interacting theories in four dimensions is still  unsettled. (On the other hand, while there has been no cure, QFT has not been 
killed by the discovery of an internal contradiction in its fundamental assumptions.)

There are two basic flavours of axiomatic QFT: the Wightman framework,\cite{StreaterWightman}
which retains the idea of a quantum field as a key building block of the theory, 
and the more radical Haag--Kastler--Araki framework of algebraic QFT (AQFT) or \emph{local quantum 
physics},\cite{Haag} in which the focus is on algebras of local observables, while fields
enter as secondary and less intrinsic elements. The motivation of the algebraic approach
is to remain close to operational ideas of what can be measured locally, simultaneously avoiding an over-reliance on classical field theory (which, after all, should emerge as a limit of QFT, rather than being taken as its foundation). Locally covariant QFT is a natural generalisation of AQFT, but retains a natural place for quantum fields.

Aside from general structural results applying to wide classes of QFTs,  axiomatic QFT also provides a deepened and better founded conceptual framework, often allied with powerful mathematical tools. In turn, this can lead to new developments, such as the formalism of perturbative algebraic QFT (pAQFT)
that has put perturbative QFT on a rigorous basis, even in curved spacetime and even
for gauge theories including gravity
(see Refs.~\refcite{BrFr2000,Ho&Wa01,Ho&Wa02,FreRej_BVqft:2012} and Rejzner's contribution to these Proceedings).  A survey of the present status of AQFT, in both flat and curved spacetimes, can be found in the edited collection Ref.~\refcite{AdvAQFT}.

\section{Locally Covariant QFT in CST}\label{sec:LCQFT}

Let us set out the general structure of locally covariant QFT. General references for
this section are the original paper,\cite{BrFrVe03} and an extensive recent review.\cite{FewVerch_aqftincst:2015}

\subsection{Locally covariant theories}

Fix a spacetime dimension $n\ge 2$. The spacetime backgrounds that we will study,
and which we will call \emph{globally hyperbolic spacetimes}, consist of tuples $\Mb  =(\Mc,g,\ogth,\tgth)$, where $\Mc$ is a smooth manifold with a Lorentzian metric $g$ (signature
$+-\cdots-$), an orientation $\ogth$, and time-orientation $\tgth$. Here, we allow $\Mc$ to have finitely many connected components, while $\ogth\subset\Omega^n(\Mc)$ is one of the components of the nowhere-vanishing smooth $n$-forms on $\Mc$ and $\tgth\subset\Omega^1(\Mc)$ is one of the components of the nowhere-vanishing smooth $1$-forms that are timelike with respect to $g$. We 
restrict to those spacetimes that are globally hyperbolic with respect to the given metric and time-orientation.

The first element of the algebraic formulation is the assignment of a $*$-algebra $\Af(\Mb)$, with a unit $\II_{\Af(\Mb)}$, to each $\Mb$ of this type. The self-adjoint elements of $\Af(\Mb)$ are to represent observables of the given theory on spacetime $\Mb$.  
A simple example is given by the real scalar field, obeying the Klein--Gordon equation
\begin{equation}
P_\Mb\phi:=(\Box_\Mb +m^2)\phi = 0.
\end{equation}
As $\Mb$ is globally hyperbolic, there are advanced ($-$) and retarded $(+)$ Green operators $E_\Mb^\pm$  for the operator $P_\Mb$ so that, for any smooth compactly supported function $f\in\CoinX{\Mb}$, $\phi^\pm = E^\pm_\Mb f$ solves the inhomogeneous
equation $P_\Mb \phi^\pm=f$ with the support of $\phi^\pm$ lying in the causal future ($+$)
or past $(-)$ of the support of $f$. Then the $*$-algebra $\Af(\Mb)$ is
defined to have a set of generators $\{\Phi_\Mb(f):f\in\CoinX{\Mb}\}$ labelled by smooth
compactly supported functions, a unit $\II_{\Af(\Mb)}$, and relations given
by
\begin{itemize}
\item linearity of $f\mapsto \Phi_\Mb(f)$
\item hermiticity: $\Phi_\Mb(f)^*=\Phi_\Mb(\overline{f})$ for all $f\in\CoinX{\Mb}$
\item field equation: $\Phi_\Mb(P_\Mb f)=0$  for all $f\in\CoinX{\Mb}$
\item commutation relations: $[\Phi_\Mb(f_1),\Phi_\Mb(f_2)]=iE_\Mb(f_1,f_2)\II_{\Af(\Mb)}$
 for all $f_1,f_2\in\CoinX{\Mb}$.
\end{itemize}
Here we have written 
\begin{equation}
E_\Mb(f_1,f_2)=\int_\Mb f_1(p) (E_\Mb f_2)(p)\dvol_\Mb(p),
\end{equation}
where $E_\Mb = E_\Mb^--E_\Mb^+$.

The specification of the algebra on each spacetime is only one part of the structure. 
An important aspect is the ability to compare the algebras on different spacetimes. 
This can be done by considering smooth maps $\psi:\Mb_1\to\Mb_2$ that are isometric, respect
orientation and time-orientation: if $\Mb_i=(\Mc_i,g_i,\ogth_i,\tgth_i)$ ($i=1,2$),
we require $g_1=\psi^*g_2$, $\ogth_1=\psi^*\ogth_2$, $\tgth_1=\psi^*\tgth_2$. 
Furthermore, $\psi$ is required to have a causally convex image in $\Mb_2$, 
thus ensuring that no causal links exist in the image that are not
already present in the original spacetime. A map $\psi$ obeying these conditions
will be called a \emph{hyperbolic embedding}. As a requirement of locality,
each hyperbolic embedding of $\Mb_1$ in $\Mb_2$ should provide an embedding of the physical content of our theory on $\Mb_1$ within that on $\Mb_2$,
represented mathematically by a unit-preserving $*$-homomorphism $\Af(\psi):\Af(\Mb_1)\to
\Af(\Mb_2)$. We demand that $\Af(\psi)$ is an injection, so that no observables are
lost in passing from a small spacetime to a larger one in which it is embedded.\footnote{This
is too stringent in some contexts where `topological observables' appear but
we set these to the side for now in the interests of a clean axiomatic framework.}
Moreover, we make the natural requirements that, if a trivial embedding is made, the algebraic embedding should be likewise trivial, and that the composition of maps arising from successive embeddings should 
agree with that of the composition of embeddings:
\begin{equation}
\Af(\id_\Mb)=\id_{\Af(\Mb)},\qquad \Af(\psi\circ\varphi)=\Af(\psi)\circ\Af(\varphi),
\end{equation}
where the second equation holds for all pairs of composable maps between spacetimes
in our class. These various requirements can be summarised by a single mathematical
assumption:\footnote{See Ref.~\refcite{MacLane} for a general reference on category theory. For the purpose of this review, the reader will not go too far wrong by thinking of a category as consisting of \emph{objects} that are `sets with structure' and \emph{morphisms} that are `structure preserving maps'. Examples include the category of topological spaces with continuous maps as morphisms, or groups with homomorphisms, as well as $\Loc$ and $\Alg$ described here. A functor between two categories maps objects and morphisms in the first to objects and morphisms in the second in a coherent way; for example, a homology functor
maps topological spaces to the appropriate homology group and continuous maps between the topological spaces to group homomorphisms between the homology groups.}
\begin{axiom}[Local covariance]\label{ax:LC}
A theory is a covariant functor $\Af:\Loc\to\Alg$, where
$\Loc$ is the category whose objects are globally hyperbolic spacetimes, and
whose morphisms are hyperbolic embeddings, while $\Alg$ is the category
of unital $*$-algebras with injective, unit-preserving $*$-homomorphisms
as morphisms.\footnote{It is of course possible to change the category $\Alg$ for, e.g., 
the category of $C^*$-algebras. Alternatively theories other than QFT can be set into
a locally covariant context by an appropriate choice of target category, e.g., 
that of (pre)symplectic spaces for classical linear field theories.}
\end{axiom}
In the context of the Klein--Gordon theory, these morphisms are easily described:
if $\psi:\Mb\to\Nb$ then $\Af(\psi)$ maps the generators of $\Af(\Mb)$
into those of $\Af(\Nb)$ by
\begin{equation}\label{eq:field}
\Af(\psi)\Phi_\Mb(f) = \Phi_\Nb(\psi_* f)
\end{equation}
for all $f\in\CoinX{\Mb}$. Here, $\psi_*$ denotes the push-forward, so that
$\psi_*f$ agrees with $f\circ \psi^{-1}$ on $\psi(\Mb)$ and vanishes elsewhere. 
The action of $\Af(\psi)$ on all other elements of $\Af(\Mb)$ is fixed by the 
requirement that it be a $*$-homomorphism obeying $\Af(\psi)\II_{\Af(\Mb)}=\II_{\Af(\Nb)}$. That this can be done consistently is a consequence of the theory of the Klein--Gordon equation
on globally hyperbolic spacetimes and the relations in $\Af(\Mb)$ and $\Af(\Nb)$. 
For example,  
$\Af([\Phi_\Mb(f_1),\Phi_\Mb(f_2)])$ can be written as either of
$\Af(iE_\Mb(f_1,f_2)\II_{\Af(\Mb)})= iE_\Mb(f_1,f_2)\II_{\Af(\Nb)}$ or
$[ \Phi_\Nb(\psi_* f_1),\Phi_\Nb(\psi_* f_2)] =iE_\Nb(\psi_*f_1,\psi_* f_2)\II_{\Af(\Nb)}$
and consistency is assured because $E_\Nb(\psi_*f_1,\psi_* f_2)=E_\Mb(f_1,f_2)$. 
It is less obvious that the resulting map is injective: this follows because $\Af(\Mb)$ is known to be simple, so the kernel of $\Af(\psi)$ is either
trivial or equals $\Af(\Mb)$, and the latter is impossible because $\Af(\psi)\II_{\Af(\Mb)}=\II_{\Af(\Nb)}\neq 0$.

Other free bosonic models have been formulated as functors from $\Loc$ to $\Alg$,
including the Proca and (with some subtleties) Maxwell fields.\cite{DappLang:2012,SandDappHack:2014,FewLang:2014a} This includes 
examples (self-dual gauge fields) which are not formulated by reference to a
classical Lagrangian.\cite{BecBenSchSza:2015}
To incorporate theories with spin, one can either generalise $\Loc$ to a category
of spin manifolds,\cite{Verch01,Sanders_dirac:2010,Zahn_Dirac:2014} or -- as  in Section~\ref{sec:spinstats} 
-- use coframed manifolds (see Ref.~\refcite{FergusonPhD} for yet another approach).

\subsection{Comparison of theories and locally covariant fields}\label{sec:nat}

Assumption~\ref{ax:LC} has two main strengths: first, it has packaged many individual
assumptions into one statement; second, it allows us to discuss a theory as a 
single mathematical object, rather than viewing it through its instantiations on 
each spacetime separately. Category theory provides a language for this discussion
and a number of standard categorical ideas find uses in locally covariant quantum field theory. 
A particularly important example is the notion of a \emph{natural transformation} between
functors (denoted by a dotted arrow $\nto$),
which has two main uses in locally covariant QFT. The first of these concerns
relations between theories: 
\begin{definition} \label{def:nat}
Let $\Af$ and $\Bf$ be locally covariant theories (functors from $\Loc$ to $\Alg$). Any natural transformation $\eta:\Af\nto\Bf$ defines an embedding of $\Af$
as a subtheory of $\Bf$. If $\eta$ is a natural isomorphism, then it determines
a physical equivalence of the theories $\Af$ and $\Bf$.
\end{definition}
Here, a natural transformation $\eta:\Af\nto\Bf$ is a collection $(\eta_\Mb)_{\Mb\in\Loc}$ 
of morphisms $\eta_\Mb:\Af(\Mb)\to\Bf(\Mb)$ such that, for every hyperbolic
embedding $\psi:\Mb\to\Nb$, one has
\begin{equation}
\eta_{\Nb}\Af(\psi) = \Bf(\psi) \eta_{\Mb}.
\end{equation}
In other words, the square in the diagram
\begin{equation*}
\begin{tikzpicture}[baseline=0 em, description/.style={fill=white,inner sep=2pt}]
\matrix (m) [ampersand replacement=\&,matrix of math nodes, row sep=3em,
column sep=2.5em, text height=1.5ex, text depth=0.25ex]
{\Mb \&  \Af(\Mb) \&  \Bf(\Mb) \\
\Nb \& \Af(\Nb) \&  \Bf(\Nb)\\ };
\path[->,font=\scriptsize]
(m-1-1) edge node[auto] {$ \psi $} (m-2-1)
(m-1-2) edge node[auto] {$ \eta_\Mb $} (m-1-3)
        edge node[auto] {$ \Af(\psi) $} (m-2-2)
(m-2-2) edge node[auto] {$ \eta_\Nb $} (m-2-3)
(m-1-3) edge node[auto] {$ \Bf(\psi) $} (m-2-3);
\end{tikzpicture}
\end{equation*}
commutes; $\eta$ is a natural isomorphism if each of its components $\eta_\Mb$ is
an isomorphism. 

The interpretation placed on natural transformations and isomorphisms
in Definition~\ref{def:nat} can be justified in several ways and has found various applications.\cite{BrFrVe03,FewVer:dynloc_theory,Fewster:gauge} In particular, the automorphisms of $\Af$ (the natural isomorphisms of $\Af$ to itself) form a group $\Gc$ under composition, which can be interpreted as the \emph{global gauge group} of the theory $\Af$.\cite{Fewster:gauge} A simple example
of a global gauge transformation for the Klein--Gordon theory is defined so that
$\eta_\Mb \Phi_\Mb(f)=-\Phi_\Mb(f)$ (extended as a unit-preserving $*$-homomorphism);
if one considers the theory of $n$ Klein--Gordon fields with identical mass one has
an ${\rm O}(n)$ group of orthogonal transformations on the multiplet of fields (and further
shift transformations if the mass is zero). An example of a subtheory embedding can
be given if $\Bf=\Af\otimes\Af$ consists of two identical copies of $\Af$,\footnote{Thus 
$\Bf(\Mb)=\Af(\Mb)\otimes\Af(\Mb)$ while $\Bf(\psi)=\Af(\psi)\otimes\Af(\psi)$.} and we define
$\eta:\Af\nto\Bf$ so that $\eta_\Mb A = A\otimes\II_{\Af(\Mb)}$, which can easily
be verified as natural. Klein--Gordon theories with distinct mass, or Klein--Gordon multiplets with differing numbers of fields, can be shown to be inequivalent (modulo additional mild technical conditions).\cite{BrFrVe03,FewVerch_aqftincst:2015}

A second use of natural transformations is to describe \emph{locally covariant fields}. 
For simplicity we restrict here to fields smeared with scalar, smooth compactly supported test functions.
Let $\Set$ be the category of sets, with functions as morphisms. This category
contains $\Alg$ as a subcategory: every unital $*$-algebra is, in particular, a set and
every $*$-homomorphism between such algebras is, in particular, a function.  
The assignment of the space of scalar test functions to spacetime $\Mb$ 
can be described as a functor $\Df:\Loc\to\Set$ by setting $\Df(\Mb)=\CoinX{\Mb}$ for each $\Mb$ and $\Df(\psi)=\psi_*$
for each hyperbolic embedding $\psi$. Locally covariant fields can be then identified as follows:\cite{Ho&Wa01, BrFrVe03}
\begin{definition}
Let $\Af$ be a locally covariant theory. A locally covariant field of the theory $\Af$
is a natural transformation $\Phi:\Df\nto\Af$, where we regard $\Alg$ as a subcategory of 
$\Set$.  
\end{definition}
This means precisely that the equation \eqref{eq:field} should hold for all $f\in\CoinX{\Mb}$ and
all $\psi:\Mb\to\Nb$, with $\Phi_\Mb$ now reinterpreted as the maps that form the components of 
$\Phi:\Df\nto\Af$. This gives the Klein--Gordon field its own
mathematical status. In general, the collection of all locally covariant fields forms a 
unital $*$-algebra $\Fld(\Df,\Af)$: given 
$\Phi,\Psi\in\Fld(\Df,\Af)$, and $\lambda\in\CC$, the
fields $\Phi+\lambda\Psi$, $\Phi\Psi$, $\Phi^*$ are
\begin{align}
(\Phi+\lambda\Psi)_\Mb(f) &= \Phi_\Mb(f) + \lambda\Psi_\Mb(f)  \label{eq:fld_lin}
\\ 
(\Phi\Psi)_\Mb(f) &= 
\Phi_\Mb(f)\Psi_\Mb(f), \label{eq:fld_prod} \\
(\Phi^*)_\Mb(f) &= \Phi_\Mb(f)^* \label{eq:fld_star}
\end{align}
and the unit field is $\II_\Mb(f) = \II_{\Af(\Mb)}$, for all $f\in\CoinX{\Mb}$. This
$*$-algebra then carries an action of the gauge group of the theory, so that if $\eta\in\Gc$ then  
$\eta\cdot\Phi$ is the field with components
\begin{equation}
(\eta\cdot\Phi)_\Mb(f)=\eta_\Mb\Phi_\Mb(f)
\end{equation}
for all $\Mb\in\Loc$, $f\in\CoinX{\Mb}$. Consequently, the fields appear in 
multiplets corresponding to subspaces of $\Fld(\Df,\Af)$ that are irreducible
under the action of $\Gc$. One can easily adapt the same idea to other types of smearing test functions.
As with the theories themselves, the locally covariant viewpoint on fields allows them
to be manipulated as mathematical objects in a spacetime-independent way. 

\subsection{The kinematic net}

Given a spacetime $\Mb$, one can ask what physical content can be associated with
a specific subregion $O\subset\Mb$. This can be achieved in a simple fashion in our
functorial setting, if $O$ is causally convex and open, for then we can equip
$O$ with the metric and (time)-orientation inherited from $\Mb$ and regard it
as a globally hyperbolic spacetime in its own right, to be denoted $\Mb|_O$. 
Moreover, the inclusion map of $O$ within $\Mb$ now induces a hyperbolic
embedding $\iota_{\Mb;O}:\Mb|_O\to\Mb$. Applying the functor, we obtain
an algebra $\Af(\Mb|_O)$ and a $*$-homomorphism $\Af(\iota_{\Mb;O})$
mapping $\Af(\Mb|_O)$ into $\Af(\Mb)$. 
\begin{definition} For  any nonempty, open, causally convex subset $O$ of $\Mb$,
the image of $\Af(\iota_{\Mb;O})$ will be denoted
$\Af^\kin(\Mb;O)$ and called the \emph{kinematic subalgebra} of $\Af(\Mb)$
associated with region $O$, and the assignment $O\mapsto\Af^\kin(\Mb;O)$
forms the \emph{kinematic net} of $\Af$ on $\Mb$.
\end{definition} 
The kinematic subalgebras have a number of nice properties.
If  $O_1\subset O_2$, then $\iota_{\Mb;O_1}=\iota_{\Mb;O_2}\circ
\iota_{\Mb|_{O_2};O_1}$, which implies that
$\Af(\iota_{\Mb;O_1})=\Af(\iota_{\Mb;O_2})\circ\Af
(\iota_{\Mb|_{O_2};O_1})$ and hence one has the \emph{isotony} relation 
\begin{equation}\label{eq:isotony}
\Af^\kin(\Mb;O_1)\subset \Af^\kin(\Mb;O_2).
\end{equation}
If $\psi:\Mb\to\Nb$ then, similarly, $\Af(\psi)(\Af^\kin(\Mb;O))=\Af^\kin(\Nb;\psi(O))$.
This is particularly interesting in the case of a symmetry of $\Mb$, i.e., an isomorphism $\alpha:\Mb\to\Mb$, in which case
\begin{equation}\label{eq:sym}
\Af(\alpha)(\Af^\kin(\Mb;O))=\Af^\kin(\Mb;\alpha(O)).
\end{equation}
Furthermore, the kinematic subalgebras fit well with the other structures we have introduced: if $\Phi\in\Fld(\Df,\Af)$,
and $f\in\CoinX{\Mb}$ is supported within $O$, then $\Phi_\Mb(f)\in\Af^\kin(\Mb;O)$. This
holds because $f=\Df(\iota_{\Mb;O})\hat{f}$ for $\hat{f}=\iota_{\Mb;O}^*f\in\Df(\Mb|_{O})$, and hence
\begin{equation}
\Phi_\Mb(f)=\Phi_\Mb(\Df(\iota_{\Mb;O})\hat{f}) = \Af(\iota_{\Mb;O})\Phi_{\Mb|_O}(\hat{f})
\in\Af^\kin(\Mb;O).
\end{equation}
Moreover, the gauge transformations act locally: $\eta_\Mb(\Af^\kin(\Mb;O))=\Af^\kin(\Mb;O)$ for 
any $\eta\in\Gc$, $\Mb\in\Loc$ and open, causally convex $O\subset\Mb$. 
It is noteworthy that in Minkowski space algebraic QFT, equations~\eqref{eq:isotony} 
and~\eqref{eq:sym}, together with the other properties just described, are separate assumptions about the theory; here, they are consequences of Assumption~\ref{ax:LC} and the definition of the kinematic subalgebras.

One normally makes a further assumption
\begin{axiom}[Einstein Causality] If $O_1$ and $O_2$ are open, causally convex regions of $\Mb$ 
that are spacelike separated ($O_1\cap J_\Mb(O_2)=\emptyset$) then
$\Af^\kin(\Mb;O_1)$ and $\Af^\kin(\Mb;O_2)$ are commuting subalgebras of $\Af(\Mb)$.
\end{axiom}
(This could be changed to a graded commutator if required.)

\subsection{The timeslice axiom and relative Cauchy evolution}
 
The structures introduced so far are kinematic in nature and lack any notion of 
dynamics. Describing any hyperbolic embedding $\psi:\Mb\to\Nb$ whose image 
contains a Cauchy surface of $\Nb$ as \emph{Cauchy}, the existence of a
dynamical law can be encapsulated in the following assumption. 
\begin{axiom}[Timeslice] If $\psi:\Mb\to\Nb$ is Cauchy, then $\Af(\psi):\Af(\Mb)\to\Af(\Nb)$
is an isomorphism.
\end{axiom}
This assumption means that any observable of the theory on $\Mb$ can be measured,
equivalently, within a neighbourhood of any Cauchy surface. It therefore corresponds
to an abstracted notion of a dynamical law. Note that no equation of motion has
been assumed in our general framework. 

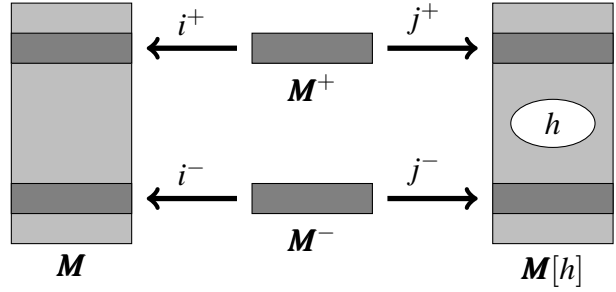
\begin{figure}[t]
\begin{center}
\begin{tikzpicture}[scale=0.8]
\definecolor{Gold}{rgb}{.93,.82,.24}
\definecolor{Orange}{rgb}{1,0.5,0}
\draw[fill=lightgray] (-4,0) -- ++(2,0) -- ++(0,4) -- ++(-2,0) -- cycle;
\draw[fill=lightgray] (4,0) -- ++(2,0) -- ++(0,4) -- ++(-2,0) -- cycle;
\draw[fill=gray] (4,3) -- ++(2,0) -- ++(0,0.5) -- ++(-2,0) -- cycle;
\draw[fill=gray] (0,3) -- ++(2,0) -- ++(0,0.5) -- ++(-2,0) -- cycle;
\draw[fill=gray] (-4,3) -- ++(2,0) -- ++(0,0.5) -- ++(-2,0) -- cycle;
\draw[fill=gray] (4,0.5) -- ++(2,0) -- ++(0,0.5) -- ++(-2,0) -- cycle;
\draw[fill=gray] (0,0.5) -- ++(2,0) -- ++(0,0.5) -- ++(-2,0) -- cycle;
\draw[fill=gray] (-4,0.5) -- ++(2,0) -- ++(0,0.5) -- ++(-2,0) -- cycle;
\draw[color=black,line width=2pt,->] (2.25,3.25) -- (3.75,3.25) node[pos=0.4,above]{$j^+$};
\draw[color=black,line width=2pt,->] (2.25,0.75) -- (3.75,0.75) node[pos=0.4,above]{$j^-$};
\draw[color=black,line width=2pt,->] (-0.25,3.25) -- (-1.75,3.25) node[pos=0.5,above]{$i^+$};
\draw[color=black,line width=2pt,->] (-0.25,0.75) -- (-1.75,0.75) node[pos=0.5,above]{$i^-$};
\draw[fill=white] (5,2) ellipse (0.7 and 0.4);
\node at (5,2) {$h$};
\node[anchor=north] at (5,0) {$\Mb[h]$};
\node[anchor=north] at (-3,0) {$\Mb$};
\node[anchor=north] at (1,3) {$\Mb^+$};
\node[anchor=north] at (1,0.5) {$\Mb^-$};
\end{tikzpicture}
\end{center}
\caption{The geometrical construction of relative Cauchy evolution}\label{fig:rce}
\end{figure}

The timeslice axiom has some striking consequences. One of the most
prominent is that the sensitivity of a theory to changes
in the metric can be described in terms of a \emph{relative Cauchy evolution}.\cite{BrFrVe03,FewVer:dynloc_theory} Fix a spacetime $\Mb=(\Mc,g,\ogth,\tgth)\in\Loc$ and let $h$ be a smooth and compactly supported metric perturbation
so that $\Mb[h]:=(\Mc,g+h,\ogth,\tgth[h])$ is also a globally hyperbolic spacetime in $\Loc$, 
where $\tgth[h]$ is the unique choice of time-orientation agreeing with $\tgth$ outside
the support of $h$. Choose any open causally convex sets $\Mc^\pm\subset\Mc$
with $\Mc^\pm\subset \Mc\setminus J^\mp_\Mb(\supp h)$. Setting $\Mb^\pm=\Mb|_{\Mc^\pm}$,
the inclusion maps of $\Mc^\pm$ into $\Mc$ induce Cauchy morphisms $\imath^\pm:\Mb^\pm\to \Mb$,
and also Cauchy morphisms $j^\pm:\Mb^\pm\to\Mb[h]$; see Fig.~\ref{fig:rce}. Each of these Cauchy morphisms
is turned into an isomorphism under the action of the functor $\Af$ and we may therefore
define an automorphism of $\Af(\Mb)$ by 
\begin{equation}
\rce_\Mb[h] = \Af(\imath^-)\circ  \Af(j^-)^{-1} \circ
\Af( j^+)\circ  \Af(\imath^+)^{-1},
\end{equation}
which is the relative Cauchy evolution induced by $h$. (One may show that $\rce_\Mb[h]$ is
independent of the specific choices of $\Mc^\pm$.)

The relative Cauchy evolution has been computed for various theories.\cite{BrFrVe03,Sanders_dirac:2010,Benini_Masters,Ferguson:2013,FewSchenkel:2015,FewLang:2014a}  
Even more, it is possible under some circumstances to take a functional derivative 
with respect to $h$, thus inducing a derivation of $\Af(\Mb)$ that can be 
interpreted as a commutator with a stress-energy tensor.\cite{BrFrVe03}
To be specific, let $f^{ab}$ be a compactly supported rank $2$-contravariant tensor field.
Then the stress-energy tensor smeared against $f$ has the following action on $A\in\Af(\Mb)$:
\begin{equation}
[\Ts_\Mb(f), A] =  \int_\Mb f_{\mu\nu} \frac{\delta\rce_{\Mb}}{\delta g_{\mu\nu}} (A)
:= \frac{2}{i}\left.\frac{d}{ds}\rce_\Mb[h(s)] A\right|_{s=0},
\end{equation}
where $h(s)$ is a differentiable family of metric perturbations with $\dot{h}(0)^{ab}=f^{(ab)}$. 
Understood in this way, $\Ts_\Mb$ turns out to be symmetric and conserved,
and in particular models it coincides with the standard stress-energy tensor. 
This a remarkable result, because at no stage was it assumed that the theory can be derived from a classical action principle.  
If external fields are incorporated into the background, one may consider variations
of them as well, leading to other conserved currents.\cite{FewSchenkel:2015}

Another application of the relative Cauchy evolution is to provide an alternative
notion of localisation to that encoded in the kinematic net. The idea is to regard
an observable as localised in a region if it is invariant under metric changes in the region's
causal complement. This leads to a \emph{dynamical net}: when this coincides with the kinematic net, the theory is said to be \emph{dynamically local} -- see Ref.~\refcite{FewVer:dynloc_theory},
where some general consequences are developed.  Various models (including
the massive Klein--Gordon theory) have been shown to be dynamically local.\cite{FewVer:dynloc2,Ferguson:2013,FergusonPhD,FewLang:2014a,FewSchenkel:2015} 
There are exceptions, which seem to stem from broken symmetries or topological
charges; see Ref.~\refcite{FewVerch_aqftincst:2015} for discussion.

\subsection{State spaces}

In algebraic QFT, states correspond to experimental preparations, while observables correspond to physical quantities to be measured. The pairing of states and observables produces the expectation value of the measurements of the given physical quantity, subject to the given preparation. 
Technically, a state on a unital $*$-algebra $\Ac$ is a linear functional $\omega:\Ac\to\CC$, which is positive in the sense that $\omega(A^*A)\ge 0$ for all algebra elements $A$, and is normalised to the value  $\omega(\II)=1$ on the algebra unit. On any algebra $\Ac$, we denote the corresponding set of states by 
$\Ac^*_{+,1}$.  Using the well-known GNS construction, any state on a unital $*$-algebra induces
a Hilbert space representation in which expectation values are given by the standard Born rule.

Experience has shown that the full set of states includes many that do not have good physical
properties and that it is better to focus attention on a smaller class, e.g., the Hadamard
states of Klein--Gordon theory. In the locally covariant context this indicates that one should
consider a set of states $\Sf(\Mb)\subset\Af(\Mb)^*_{+,1}$ on each spacetime $\Mb$. As
measurements made in a small spacetime should also be possible within a larger one,
there should be an appropriate relation between $\Sf(\Mb)$ and $\Sf(\Nb)$ whenever
there is a hyperbolic embedding $\psi:\Mb\to\Nb$. 
\begin{definition} 
A \emph{state space} $\Sf$ for a locally covariant theory $\Af:\Loc\to\Alg$ 
is an assignment of a subset $\Sf(\Mb)\subset \Af(\Mb)^*_{+,1}$ that is closed under
convex combinations and operations induced by $\Af(\Mb)$,\footnote{That is, given states $\omega,\omega'\in\Sf(\Mb)$, each state $\lambda\omega+(1-\lambda)\omega'$ $(\lambda\in[0,1]$) also belongs to $\Sf(\Mb)$,
as also does the state $\omega_A$ given by $\omega_A(C)=\omega(A^*CA)/\omega(A^*A)$
for any $A$ such that $\omega(A^*A)>0$.} and obeys
\begin{equation}\label{eq:sts_cont}
\Af(\psi)^*\Sf(\Nb) \subset \Sf(\Mb)
\end{equation}
whenever $\psi:\Mb\to\Nb$ is a hyperbolic embedding.\footnote{One can give a more categorical definition, in which $\Sf$ is
a subfunctor of the contravariant functor assigning to each algebra $\Af(\Mb)$ its full state space.} If \eqref{eq:sts_cont} holds with 
equality for all Cauchy hyperbolic embeddings, then $\Sf$ has the \emph{timeslice property}. 
\end{definition}
As already mentioned, the Hadamard states provide an example of such a state space, 
for the Klein--Gordon theory.  

To understand the significance of the above definition, 
suppose that $A\in\Af(\Mb)$ is an observable on spacetime $\Mb$ that is hyperbolically
embedded in $\Nb$ by $\psi$. Then there is an observable on spacetime $\Nb$,  
$\Af(\psi)A$, which corresponds to $A$. To any state on  $\omega\in\Sf(\Nb)$ 
there is an expectation value $\omega(\Af(\psi)(A))$, which can be written as 
$(\Af(\psi)^*\omega)(A)$, i.e., the expectation of $A$ in the `pulled back' state
$\Af(\psi)^*\omega$ on $\Af(\Mb)$. The content of   \eqref{eq:sts_cont} is that this pulled-back state belongs to the state space $\Sf(\Mb)$ and so is a legitimate physical state on $\Mb$. 
Accordingly, the same measurement results can be obtained either in $\Mb$ or $\Nb$.

A reasonable question is whether one can find a state space consisting of a single state
in each spacetime (even dropping the requirement of closure under operations), which would amount to a choice of a preferred state. However, it can be shown that this is impossible for dynamically local theories that obey standard assumptions in Minkowski space.\cite{FewVer:dynloc_theory} This
turns long-standing folk-wisdom into a rigorous theorem.

\subsection{Applications of the locally covariant framework}

The locally covariant framework was first introduced about 15 years ago and
has already led to substantial progress in understanding both general structural
features of QFT in CST and also specific physical problems. 

Above
all, the ideas of local covariance were instrumental in completing the perturbative construction
of interacting theories in curved spacetime\cite{BrFr2000,Ho&Wa01,Ho&Wa02,Zahn_Dirac:2014}, and
extending it to include theories with local gauge symmetries\cite{Hollands:2008,FreRej_BVqft:2012}
and gravity (see Ref.~\refcite{BruFreRej_gravity} and Rejzner's contribution to these Proceedings). In the context of gravity, key issues are the identification of suitable gauge-invariant observables (see also Ref.~\refcite{Khav:2015}) and the use of relative Cauchy evolution in the discussion of background independence (see Ref.~\refcite{Zahn_bgindep:2015} for similar considerations in another context). Anomalies have also been studied.\cite{Zahn:2015} 

In addition, there have
been a number of results of a structural nature: these include Verch's proof of the
spin-statistics connection,\cite{Verch01} Sanders' results on the Reeh--Schlieder property,\cite{Sanders_ReehSchlieder} and the analysis of the superselection structure of
locally covariant theories,\cite{Br&Ru05} including the identification of topological
sectors in suitable spacetimes.\cite{BrunettiRuzzi_topsect} The fundamental 
question of whether a locally covariant theory can be said to represent the
same physics in all spacetimes has been discussed; the issue is subtle, but there
are positive results at least for the class of dynamically local theories.\cite{FewVer:dynloc_theory}
As already described, the global gauge
group has been understood at the functorial level,\cite{Fewster:gauge} along with
the intrinsic definition of the stress-energy tensor.\cite{BrFrVe03} Quite recently,
the split property\cite{Few_split:2015} (see also Ref.~\refcite{Few_DMV:2016} for a review) and modular nuclearity\cite{LecSan:2015} 
have been proved in the locally covariant framework, given suitable additional
assumptions. Extensions of the locally covariant framework towards `higher' categorical
structures are also under way.\cite{BenSchSza:2015}

Finally, locally covariant ideas have been applied to problems in the theory of Quantum Energy Inequalities  \cite{Few&Pfen06, Marecki:2006,Fewster2007} (e.g.\ to obtain a priori bounds on Casimir energy densities) and to questions in cosmology.\cite{DapFrePin2008,DegVer2010,VerchRegensburg,HackPin_chap:2015}
 The remainder of this contribution will focus on the spin-statistics connection.

\section{Spin and Statistics}\label{sec:spinstats}
\subsection{Introductory remarks}

Observed elementary particles are either
bosons of integer spin, or fermions of half-integer spin. 
Explanations of this connection between spin and statistics 
have been sought since the early days of quantum field theory \cite{Fierz:1939,Pauli:1940} and the rigorous proof of a
connection between spin and statistics was an early and major
achievement of the axiomatic Wightman framework.\cite{Burgoyne:1958, LuedersZumino:1958,StreaterWightman} Similarly, general results
have been proved in the Haag--Kastler framework.\cite{Epstein:1967,DHRiv,GuidoLongo:1995}
In this section, we will discuss how the connection can be established in the 
locally covariant framework, after suitable adaptations.

To set the scene, let us recall the spin-statistics theorem of Burgoyne\cite{Burgoyne:1958} (see also \S II.5 in Ref.~\refcite{Haag}) which concerns a Wightman theory in Minkowski
space, with Hilbert space $\HH$ and vacuum state vector $\Omega$. The universal
cover $\SL(2,\CC)$ of the proper orthochronous Lorentz group $\LL^\uparrow_+$ is
unitarily represented on $\HH$ by $S\mapsto U(S)$. 
Let $\Phi(x)$ be a component of a spin $J$ field, i.e., $\Phi$ is one of a multiplet
of fields $\Phi_\alpha$ transforming as
\begin{equation}
U(S) \Phi_\alpha(x) U(S)^{-1} = D(S^{-1})_\alpha^{\phantom{\alpha}\beta} \Phi_\beta (\pi(S)x) ,
\end{equation}
where $D$ is a spin-$J$ $\SL(2,\CC)$ representation and
\begin{equation}
\pi:\SL(2,\CC)\to \Lc^\uparrow_+, \qquad 
\sigma_\mu \pi(S)^\mu_{\phantom{\mu}\nu} = S \sigma_\nu S^*
\end{equation}
is the covering map. Burgoyne argues that the two-point function
$\ip{\Omega}{\Phi(x)\Phi^*(y)\Omega}$ can be extended analytically
and displays invariance under the \emph{complex} Lorentz group, leading
to the identity
\begin{equation}
\ip{\Omega}{\Phi(x)\Phi^*(y)\Omega} = 
(-1)^{P+2J} 
\ip{\Omega}{\Phi^*(-y)\Phi(-x)\Omega} 
\end{equation}
for spacelike separated $x,y$, where $P$ is fixed by 
\begin{equation}
\Phi(x)\Phi^*(y) = (-1)^P \Phi^*(y)\Phi(x). 
\end{equation}
In consequence, for any test function $f$, 
\begin{equation}
\|\Phi^*(f)\Omega\|^2 = (-1)^{P+2J} 
\|\Phi(Rf)\Omega\|^2 \qquad\text{where}~ (Rf)(x)=f(-x),
\end{equation}
so (except for trivial $\Phi$) the spin--statistics connection $P=2J\pmod{2}$ holds. 
A more algebraic expression of the connection is the statement that
\begin{equation}\label{eq:spinstats}
A_1 A_2 = (-1)^{P_1 P_2} A_2 A_1, \qquad\text{if}~U(-\II)A_i U(-\II)^{-1}=(-1)^{P_i}A_i,
\end{equation}
where the $A_i$ are combinations of fields smeared in regions $O_i$ at spacelike separation or, more generally, any local observables associated with these regions;
the previous statement of the spin-statistics connection is a special case of \eqref{eq:spinstats}
because
\begin{equation}
U(-\II) \Phi_\alpha(x) U(-\II)^{-1} = D(-\II)_\alpha^{\phantom{\alpha}\beta} \Phi_\beta (x) =
(-1)^{2J}\Phi_\alpha (x) .
\end{equation}   
 
A number of assumptions play crucial parts in this argument, as can be seen from
various well-known evasions of the standard spin-statistics relation. Hilbert space 
positivity evidently plays a decisive role, and indeed ghosts provide examples of anticommuting
integer spin fields. Nonrelativistic fields, or even relativistic fields in infinite-dimensional
multiplet, can also violate the spin-statistics connection, so the properties of $U(S)$ are crucial. The analytic continuation argument depends on energy positivity 
(bosonic statistics may be imposed on a Dirac field at the cost of sacrificing positivity of the Hamiltonian\cite{Pauli:1940}) and the ability to obtain a spacetime reflection 
symmetry $x\mapsto -x$ in the identity connected component of the complex Lorentz group. 

General curved spacetimes have no geometrical symmetries, no global notion of energy positivity and the $n$-point functions of typical states of interest are not expected to have analytic  extensions. Burgoyne's proof, like all the other general proofs mentioned above, therefore has no traction in curved spacetime and there is no obvious way to repair it. Indeed, for many years, work on the spin-statistics
connection in curved spacetimes was restricted
to demonstrations that free models become inconsistent
on general spacetimes if equipped with the wrong
statistics (e.g., imposing anticommutation relations
on a scalar field)\cite{Wald_Smatrix:1979,ParkerWang:1989} 
unless some other 
property such as positivity is sacrificed.\cite{HiguchiParkerWang:1990}

The breakthrough was made by Verch,\cite{Verch01}
who established a general spin-statistics theorem for 
theories defined on each spacetime by a single field which, in particular,
obeys Wightman axioms in Minkowski space. Together with Ref.~\refcite{Ho&Wa02}, this paper was responsible for laying down many of the 
foundations of the locally covariant framework for QFT in curved spacetimes  
described in Section~\ref{sec:LCQFT}. 
Verch's assumptions allow certain properties of the theory on 
one spacetime to be deduced from its properties on another,
provided the spacetimes are suitably related by restrictions
or deformations of the metric. In particular, the spin-statistics
connection is proved by noting that, if it were violated in any one spacetime, it 
would then be violated in Minkowski space, contradicting the classic spin-statistics theorem. 

There are nonetheless some good reasons to revisit the spin-statistics connection.  
First, as a matter of principle, one hopes to gain a
better understanding of why spin is the correct concept
to investigate in curved spacetime, given the lack of
the rotational symmetries that are so closely bound up
with the description of spin in Minkowski space.
Second, Ref.~\refcite{Verch01} described spinor fields 
as sections of various bundles associated to the spin bundle. 
While this is conventional wisdom in QFT in CST,  it has the effect of
basing the discussion on geometric structures that are, in part, unobservable. 
This is unproblematic as long as the goal is to
understand particular models such as
the Dirac field; however, in order to understand the spin-statistics
connection for general theories one needs a more fundamental
starting point that avoids the insertion of spin by hand. 
Third, Ref.~\refcite{Verch01} confined itself to theories in which the algebra
in each spacetime is generated by a single field, and the argument
is indirect in parts.  

This section outlines a new and
operationally well-motivated perspective on the spin-statistics connection 
in which spin emerges as a natural concept in curved spacetimes,
and which leads to a more general and direct proof of
the connection. In particular, there is no longer any need to
describe the theory in terms of one or more fields. 
Full details will appear shortly;\cite{Few_spinstats} see also Ref.~\refcite{Few_Regensburg:2015} for a summary.

The main new ideas are contained in a generalization of locally covariant QFT based on a category of spacetimes with global coframes (i.e.,  a `rods and clocks' account of  spacetime measurements). As in Ref.~\refcite{Verch01} the goal is to prove that a spin-statistics connection in curved spacetime is implied by the standard results holding in Minkowski space; however, the proof becomes quite streamlined in the new formulation and the overall result can be formulated at a functorial level. 

Putting frames at the centre of the approach introduces redundancies because two coframes related by a global proper Lorentz transformation ought not to be physically distinguishable. A key part of the formalism involves tracking these redundancies, which leads naturally from an operational starting-point to a description that allows for spin.  The functorial setting of locally covariant quantum field theory provides various tools that make this possible, but in a way that connects with traditional understandings of spin in, for example,  Wightman field theory. 

\subsection{Locally covariant theories on coframed spacetimes}


Our new spacetime category consists of coframed globally hyperbolic spacetimes, denoted $\FLoc$. 
The objects are pairs $\Mbb=(\Mc,e)$, where $\Mc$ is a smooth manifold
of dimension $n$ on which $e=(e^\nu)_{\nu=0}^{n-1}$ is a global coframe, such that
\begin{equation}
\Lf(\Mc,e):= (\Mc, \eta_{\mu\nu} e^\mu e^\nu, [e^0], [e^0\wedge\cdots\wedge e^{n-1}])
\end{equation} 
defines a spacetime in the category $\Loc$. A morphism $\psi:(\Mc,e)\to (\Mc',e')$ in $\FLoc$ is, by definition, a smooth $\psi:\Mc\to\Mc'$ that induces a $\Loc$-morphism from $\Lf(\Mc,e)$ to $\Lf(\Mc',e')$ and obeys $\psi^*e' = e$.  
Nothing is lost, relative to the framework of Section~\ref{sec:LCQFT}, because each theory $\Af:\Loc\to\Phys$ induces a theory
$\Af\circ\Lf:\FLoc\to\Phys$, and in four dimensions, every theory described using spin bundles has a similar reformulation on $\FLoc$.\footnote{Note that all orientable four-dimensional globally hyperbolic manifolds admit global coframings.} On the
other hand, as the geometry does not fix a choice of frame, the
use of $\FLoc$ introduces redundancies that must be tracked. 
 
An important point is that these redundancies can be expressed functorially. To each $\Lambda\in \Lc^\uparrow_+$, there is a functor
$\Tf(\Lambda):\FLoc\to\FLoc$, 
\begin{equation}
\Tf(\Lambda)(\Mc,e) = (\Mc,\Lambda e) \qquad (\Lambda e)^\mu=
\Lambda^\mu_{\phantom{\mu}\nu} e^\nu
\end{equation}
with action on morphisms uniquely fixed so that $\Lf\circ\Tf(\Lambda)(\psi)=\Lf(\psi)$. In other words, $\Tf(\Lambda)$ is a
rigid rotation of the frames in all spacetimes. Every theory $\Af:\FLoc\to\Phys$ now induces a family of theories $\Af\circ\Tf(\Lambda)$ labelled by $\Lambda\in\Lc^\uparrow_+$. 
Our fundamental assumption is that all these theories should be physically equivalent, with such equivalences encoded by natural isomorphisms as explained in Section~\ref{sec:nat}.  Thus we assume that to each $\Lambda\in\Lc^\uparrow_+$, there exists an equivalence
$\eta(\Lambda): \Af\nto\Af\circ\Tf(\Lambda)$; for convenience we also assume that the $\eta(\Lambda)$
`commute' with the action of the global gauge group $\Gc$.
  
It is remarkable that this assumption (without further specification of the $\eta(\Lambda)$'s) already yields a number of consequences. First,  consider successive transformations $\Lambda$ and $\Lambda'$. 
There are two ways of comparing $\Af$ with $\Af\circ\Tf(\Lambda'\Lambda)$: directly using $\eta(\Lambda'\Lambda)$, 
or in two steps, using $\eta(\Lambda)$ followed by $\Lambda^*\eta(\Lambda')$, the equivalence defined by $(\Lambda^*\eta(\Lambda'))_\Mbb=\eta(\Lambda')_{\Tf(\Lambda)(\Mb)}$. The comparison between them, i.e., the extent to which the diagram
\begin{equation*}
\begin{tikzpicture}[baseline=0 em, description/.style={fill=white,inner sep=2pt}]
\matrix (m) [ampersand replacement=\&,matrix of math nodes, row sep=3em,
column sep=2.5em, text height=1.5ex, text depth=0.25ex]
{ \Af \&  \Af\circ\Tf(\Lambda) \\
        \&  \Af\circ\Tf(\Lambda'\Lambda)\\ };
\path[->,font=\scriptsize]
(m-1-1) edge node[auto] {$ \eta(\Lambda) $} (m-1-2)
        edge node[below,sloped] {$ \eta(\Lambda'\Lambda) $} (m-2-2)
(m-1-2) edge node[auto] {$ \Lambda^*\eta(\Lambda') $} (m-2-2);
\end{tikzpicture}
\end{equation*}
fails to commute, is measured by a \emph{$2$-cocycle} $\xi(\Lambda,\Lambda')$ of $\Lc^\uparrow_+$ in $\Zc(\Gc)$, the centre of $\Gc$. Importantly, while there is freedom in choosing the $\eta(\Lambda)$'s, 
$\xi$ is unique up to cohomological equivalence, so \emph{each
theory $\Af$ on $\FLoc$ determines a canonical cohomology class $[\xi]\in H^2(\Lc^\uparrow_+;\Zc(\Gc))$.} Under some circumstances, $[\xi]$ is trivial: e.g., if $\Zc(\Gc)$ is trivial or $\Af$ is induced from a theory on $\Loc$ (as we may take $\eta(\Lambda)=\id$ for every
$\Lambda$).

As discussed in Section~\ref{sec:LCQFT}, the \emph{scalar} (one-component) fields of the theory
form a $*$-algebra $\Fld(\Df,\Af)$. This algebra now carries actions of both the gauge group and the Lorentz group:
\begin{align*}
(\alpha\cdot\Phi)_{(\Mc,e)} (f) &= \alpha_{(\Mc,e)}\Phi_{(\Mc,e)}(f)
& (\alpha\in\Gc)\\
(\Lambda\star \Phi)_{(\Mc,\Lambda e)}(f) &= \eta(\Lambda)_{(\Mc,e)}\Phi_{(\Mc,e)}(f) &(\Lambda\in\Lc^\uparrow_+).
\end{align*}
These actions commute, and one finds  
\begin{equation}
(\Lambda'\Lambda)\star\Phi= \xi(\Lambda',\Lambda)\cdot
(\Lambda'\star(\Lambda\star\Phi)),
\end{equation}
from which it follows that irreducible subspaces of $\Fld(\Df,\Af)$ under the
action of $\Lc^\uparrow_+\times \Gc$ carry multiplier representations of $\Lc^\uparrow_+$,  determined
by $\xi$. This `rediscovers' the classification of fields according to  
representations of the universal cover of $\Lc^\uparrow_+$.  
 
So far we have avoided specifying the $\eta(\Lambda)$. However the
dynamics of the theory suggests a way to construct them. Consider the
spacetimes $(\Mc,e)$ and $(\Mc,\Lambda e)$. Let $\tilde{\Lambda}\in C^\infty(\Mc,\Lc^\uparrow_+)$ agree with $\Lambda$ (resp., the identity)
everywhere to the future (resp., past) of suitably chosen Cauchy 
surfaces. Thus it is locally constant outside a time-compact set.
Then the spacetime $(\Mc,\tilde{\Lambda}e)$ interpolates
between $(\Mc,e)$ (with which it agrees sufficiently to the past)
and $(\Mc,\Lambda e)$ (with which it agrees sufficiently to the future).
Assuming the theory obeys the timeslice axiom, we may obtain
in this way an isomorphism between $\Af(\Mc,e)$ and $\Af(\Mc,\Lambda e)$
(cf.\ the construction of relative Cauchy evolution in Section~\ref{sec:LCQFT}).
Without further assumptions, this isomorphism could depend on the details
of $\tilde{\Lambda}$. However, it is reasonable to assume that 
frame rotations that are trivial outside
a time compact set and are homotopically trivial within this class
induce a trivial relative Cauchy evolution. In this case, 
the isomorphism from $\Af(\Mc,e)$ to $\Af(\Mc,\Lambda e)$ depends on $\tilde{\Lambda}$ only via
its homotopy class among $C^\infty(\Mc,\Lc^\uparrow_+)$ maps
that are locally constant outside time-compact sets. 
The upshot is that each $S$ in the universal cover 
of $\Lc^\uparrow_+$ induces an isomorphism 
\begin{equation}
\zeta_{(\Mc,e)}(S): 
\Af(\Mc,e)\longrightarrow \Af(\Mc,\pi(S) e).
\end{equation}
Provided that these form the components of a natural isomorphism
$\zeta(S):\Af\nto \Af\circ\Tf(\pi(S))$ (which holds subject to an additivity assumption) we may obtain our required family of equivalences by
$\eta(\Lambda)=\zeta(S_\Lambda)$, where $S_\Lambda$ is any lift of $\Lambda$
to the universal cover. 

Applying these structures in $n=4$ dimensions, one finds that $\zeta(-\II)$ is a global gauge transformation $\zeta(-\II) \in\Gc$ obeying $\zeta(-\II)^2 = \zeta(\II)= \id$. Moreover, in Minkowski spacetime $\Mbb_0=(\RR^4,(dx^\mu)_{\mu=0..3})$, if $\SL(2,\CC)$ is unitarily implemented by 
$S\mapsto U(S)$ then $\zeta(S)_{\Tf(\pi(S)^{-1})(\Mbb_0)}\circ\Af(\psi_{\pi(S)})$ is
implemented by $\ad_{U(S)}$. 
This shows how the various aspects of the standard Minkowski transformation under $U(S)$ are implemented in the locally covariant framework: the active transformation of points is achieved
by the functor $\Af(\psi_{\pi(S)})$ while the passive relabelling of field components is 
done by $\zeta(S)$. In particular, in the special case $S=-\II$, $\zeta(-\II)_{\Mbb_0}$ is implemented by the adjoint action of $U(-\II)$, i.e., the $2\pi$ rotation,
and can be termed the \emph{univalence automorphism}.

\subsection{The spin-statistics connection in four dimensions} 

We first need a definition of `statistics' at the functorial level. 
An involutory global gauge transformation $\gamma\in\Gc$, $\gamma^2=\id$ will be said to \emph{grade statistics in $\Mbb$} if
\begin{equation}
A_1 A_2 = (-1)^{\sigma_1\sigma_2} A_2 A_1
\end{equation}
holds for all local operators $A_i\in\Af^\kin(\Mbb;O_i)$, where $O_1$ and $O_2$ are 
spacelike separated, and obeying  $\gamma_{\Mbb} A_i=(-1)^{\sigma_i}A_i$. From this point of view, 
the standard spin-statistics connection precisely asserts that $\zeta(-\II)$ grades statistics in Minkowski space $\Mbb_0$. What can be proved is that, if such a $\gamma$ grades statistics in $\Mbb_0$, then
it does so in every spacetime of $\FLoc$, an argument that depends critically on the timeslice
property. Now, if the theory obeys the standard spin-statistics
connection in Minkowski space -- for example, if $\Af(\Mbb_0)$ can be identified with a
Wightman theory -- then $\zeta(-\II)$ must grade statistics in  
every $\Mbb\in\FLoc$.\cite{Few_spinstats,Few_Regensburg:2015}  
What this means is that the statistics are directly related to the $2\pi$-rotation of frames
in every spacetime, and indeed, the statement can be made in a spacetime-independent fashion 
that $\zeta(-\II)$ grades statistics for $\Af$. While the proof is indirect, because one argues from the connection in Minkowski space rather than proving it afresh in each spacetime, this of course
does not detract from the worth of the statement.

As mentioned at the start of this section, 
the spin-statistics connection is rather subtle, and even Feynman was forced onto the defensive:
\begin{quote}
We apologize for the fact that we cannot
give you an elementary explanation. An explanation has been
worked out by Pauli from complicated arguments of quantum
field theory and relativity.  ...[W]e have not been able to find a way
of reproducing his arguments on an elementary level. ...
The explanation is deep down in relativistic quantum mechanics. 
This
probably means that we do not have a complete understanding
of the fundamental principle involved. 
[RP Feynman, Lectures on Physics III (\S 4.1)]\cite{FeynmanIII}
\end{quote} 
For the moment, I have to add my own apologies for the lack of a direct
proof. However, that one can prove structural results of quantum field
theory in curved spacetime at all is a notable achievement, and indicates
the power of the locally covariant framework.

\section*{Acknowledgments}
I am grateful to the organisers of the QF2 session for arranging partial financial support under the ERC Advanced Grant ``Operator Algebras and Conformal Field Theory'' (PI Roberto Longo). 
 
%

\end{document}